\documentclass[12pt]{iopart}
\usepackage{epsfig}
\usepackage{graphicx}

\newcommand{\be}{\begin{equation}}
\newcommand{\ee}{\end{equation}}
\newcommand{\bd}{\begin{displaymath}}
\newcommand{\ed}{\end{displaymath}}

\newcommand{\bra}{\langle}
\newcommand{\ket}{\rangle}
\newcommand{\order}{{\cal O}}

\newcommand{\bsigma}{{\mbox{\boldmath $\sigma$}}}

\newcommand{\bh}{\ensuremath{\mathbf{h}}}

\newcommand{\bu}{\ensuremath{\mathbf{u}}}
\newcommand{\bv}{\ensuremath{\mathbf{v}}}

\newcommand{\bx}{\ensuremath{\mathbf{x}}}

\newcommand{\bT}{\ensuremath{\mathbf{T}}}

\newcommand{\bH}{\ensuremath{\mathbf{H}}}
\newcommand{\cM}{\ensuremath{\mathcal{M}}}
\newcommand{\cN}{\ensuremath{\mathcal{N}}}
\newcommand{\R}{{\rm I\!R}}
\usepackage{epsfig}
\usepackage{graphicx}
\usepackage{eufrak}

\newcommand{\btau}{\mbox{\boldmath $\tau$}}

\newcommand{\tPhi}{{\tilde{\Phi}}}
\newcommand{\tPsi}{{\tilde{\Psi}}}
\newcommand{\tW}{{\tilde{W}}}
\newcommand{\arctanh}{ ~{\rm arctanh} }

\begin{document}

\title{Replica symmetry breaking in the `small world' spin glass}
\author{B Wemmenhove$^1$, T Nikoletopoulos$^2$ and J P L Hatchett$^2$}
\address{$^1$ ~Institute for
Theoretical Physics, University of Amsterdam, Valckenierstraat 65,
1018 XE Amsterdam, The Netherlands}
\address{$^2$ ~ Department of Mathematics, King's College London, The Strand,
London WC2R 2LS, United Kingdom}

\begin{abstract}
We apply the cavity method to a spin glass model on a `small world'
lattice, a random bond graph 
super-imposed upon a 1-dimensional ferromagnetic ring.
We show the correspondence with a replicated
transfer matrix approach, up to the level of one step replica
symmetry breaking (1RSB). Using the scheme developed by M\'{e}zard \&
Parisi for the Bethe lattice, we evaluate observables for a model with
fixed connectivity and $\pm J$ long range bonds. Our results
agree with numerical simulations significantly better than the replica
symmetric (RS) theory.  
\end{abstract}


\ead{\mailto{wemmenho@science.uva.nl},
\mailto{theodore@mth.kcl.ac.uk},\mailto{ hatchett@mth.kcl.ac.uk}}

\section{Introduction}
Small world networks \cite{watts-strogarz98} have become
popular models in a variety of areas in physics and
mathematics \cite{monasson99,barrat-weigt00,gitterman00,kim01,herrero02}, 
as their architecture mimics that found in many real-world situations. 
Social networks, the world wide web, infra-structure networks, as well as
networks in biology (neural networks, protein regulatory networks, etc.)
are all thought to exhibit the `small world effect'. This means that 
the average minimal path length in such a network is 
relatively small compared with the topological distance, due to the
presence of long range `short cuts'. This natural distance depends on 
the backbone of the graph, typically a one-dimensional ring, as in  
\cite{Ni04}, where a spin glass on a small world network 
was studied. It was argued that this might be a 
model for RKKY type interactions in metallic spin-glasses.
The short range interactions were
uniformly ferromagnetic while the long range bonds were taken to be random. 
The randomness in the structure of the long range graph and the value
of the long range  
interactions constitutes quenched disorder and was treated using
replicated transfer matrices for the one dimensional ring
\cite{nik-coolen04}. Within this formalism, an Ising spin   
model was solved at a replica symmetric (RS) level of approximation,
where the calculation 
of observables away from the paramagnetic phase required population
dynamics algorithms. Although the RS solution 
was a good first approach, it was anticipated that there
would certainly be parameter regions for which the RS
theory would not be sufficient and replica symmetry breaking would
have to be taken into account, e.g. for vanishing values of the short
range interactions 
where one recovers a spin glass on a random graph.

In this paper we demonstrate how the cavity method can be applied to `small world'
models and extend the replica formalism solution of \cite{Ni04} to the
one step replica symmetry breaking (1RSB) level. 
After introducing the model,
we present the cavity method in RS approximation, for our 
`small world' lattice and show the equivalence with the
replica method. We then proceed with the 1RSB cavity calculation along the lines of 
\cite{MePa01}, present the appropriate extension of the replica formalism
(in terms of replicated transfer matrices) and show the correspondence between the
two methods at the 1RSB level.
Finally, we apply our theory to a point in phase space
where the replica symmetric theory predicts zero magnetization but a
non reentrant conjecture \cite{PaTo} suggests that the magnetization
order parameter should be nonzero.
We find that this point displays a finite magnetization at the 1RSB
level. Support for this finding is provided with simulation results.

\section{Model definitions}
Our model describes $N$ interacting Ising spins $\sigma_i \in \{-1, 1\}$, 
$i \in \{1,\ldots, N\}$, in thermal equilibrium at inverse
temperature $\beta=1/T$, on a lattice consisting of a 1-dimensional ring 
to which we add random long range bonds at each site. The number of long 
range bonds $k_i$ at site $i$ is distributed identically for each site
according to the degree distribution $p(k)$, the mean of which we
denote by $\bra k \ket$. For a Poisson distribution $p(k)$ we 
recover the model of \cite{Ni04}. The Hamiltonian of the system is given by
\begin{equation}
H = -J_0 \sum_i \sigma_i \sigma_{i+1} - \frac{1}{\langle k \rangle}
\sum_{i<j} J_{ij} c_{ij} \sigma_i \sigma_j
\end{equation}
with $\sigma_{N+1}\equiv \sigma_1$ and 
where $J_{ij}$ and $c_{ij}$ are quenched random variables.
The long range interactions $J_{ij} \in \R$ are independently drawn 
from a distribution $p_{J}(J_{ij})$, while the symmetric ($c_{ji}=c_{ij}$)
dilution variables 
$c_{ij} \in \{0, 1\}$ are random, but obey the constraints
\begin{equation}
\sum_{j} c_{ij} = k_i ~~~~~~~~~~ \forall i
\end{equation}
Throughout we assume that $J_0 \geq 0$, which is appropriate for
investigating RKKY type interactions. It is expected 
that negative short range interactions could
lead to a variety of complex phenomena such as
first order phase transitions between multiple locally stable states or
distributions of fields with fractal support (see e.g
\cite{skantzos-coolen00}), which 
would certainly be of interest from the point of view of complex systems but go
beyond the scope of this paper.

\section{The distribution of cavity fields at the RS level}
Throughout the following we apply a modification of the cavity method used
to study the Bethe lattice in 
\cite{MePa01} and \cite{MePa03} which, for the sake of compactness,
 we refer to for
a detailed discussion of the method. To simplify the discussion we
initially consider the case $p(k^\prime) = \delta_{k^\prime,k}$  i.e. the 
long range connectivity is fixed to the value
$k$. Later we generalize to ensembles of graphs with arbitrary $p(k^\prime)$.

\subsection{Distribution of cavity fields for the Bethe lattice}
The idea behind the cavity method for the Bethe lattice at the level of 
replica symmetry is the assumption that the mutual statistical
dependence of a set of spins,  $\{\sigma_1, \ldots, \sigma_{k-1}\}$,
which all
interact with a single common spin $\sigma_0$,
is due only to the spin $\sigma_0$. Thus, \emph{a priori}, 
their individual state statistics would be identical in the 
absence of that spin. The individual spin probabilities, when their link to 
$\sigma_0$ is removed, (spins which miss one link 
are called `cavity spins') can consequently be 
characterised by independent identically distributed effective local fields
$\{h_1, \ldots, h_{k-1}\}$ where
$p(\sigma_{j}) \sim \rme^{\beta h_j \sigma_j}$. Self-consistent
equations for the 
distribution of effective fields $\{h_j\}$
can then be found by linking the $k-1$ cavity spins with $\sigma_0$,
which is then itself a new cavity spin.
The partition function of the new cavity spin is
\begin{eqnarray}
Z(\sigma_0)  & = &\sum_{(\sigma_1, \ldots, \sigma_{k-1})}
\exp\left[\beta\left(\sigma_0 \sum_{\ell=1}^{k-1} J_\ell \sigma_\ell 
+ \sum_{\ell=1}^{k-1} h_\ell \sigma_\ell \right)\right] 
\nonumber \\
& = & c^{-1}(\{ J_\ell\}, \{ h_\ell \}) \exp\left[\beta \sigma_0 \sum_{\ell=1}^{k-1} 
u(J_\ell,h_\ell)\right]
\label{partsum}
\end{eqnarray}
with
\begin{equation}
u(J_\ell, h_\ell) = \frac{1}{\beta} \tanh^{-1}[\tanh(\beta J_\ell)\tanh(\beta h_\ell)]
\end{equation}
and
\begin{equation}
c(\{ J_\ell \}, \{ h_\ell \}) = \prod_\ell 2 \frac{\cosh(\beta J_\ell)
  \cosh(\beta h_\ell)}{\cosh(\beta u(J_\ell, h_\ell))}
\end{equation}
The RS assumption implies $p(\sigma_{0})\sim e^{\beta h_{0}\sigma_{0}}$, therefore we identify
\begin{equation}
h_0 = \sum_{\ell=1}^{k-1} u(J_\ell, h_\ell)
\end{equation}
as the cavity field of the new cavity spin $\sigma_{0}$.
Invariance of the distribution of effective fields
under this graph operation (the linking to the new cavity spin) leads to the 
iterative equation for the distribution of cavity fields 
\begin{equation}
W(h) = \int D_{k-1}
~\delta[ h - \sum_{\ell=1}^{k-1} u(J_\ell, h_\ell)]
\end{equation}
for a Bethe lattice with connectivity $k$. We have introduced the 
shorthand (to be used throughout the paper)
$D_k = \prod_{\ell = 1}^k \rmd h_\ell \rmd J_\ell W(h_\ell)p_{J}(J_\ell)$.
\subsection{Distribution of
cavity fields for the small world lattice}
\begin{figure}[t]
\begin{picture}(200,150)
\put(25,15){\includegraphics[width=0.4\textwidth]{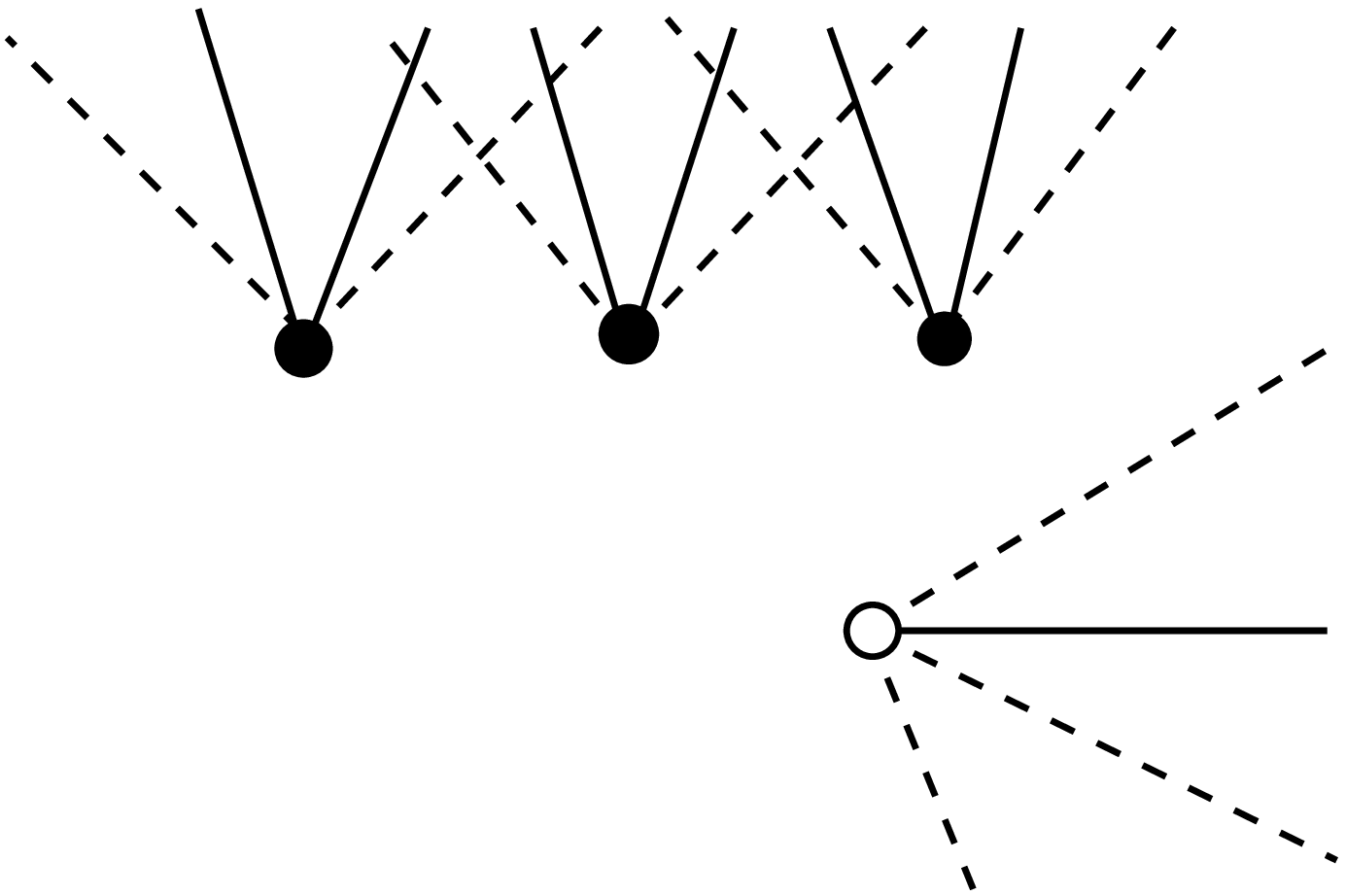}}
\put(45,85){\small $\tau_1$}
\put(88,85){\small $\tau_2$}
\put(131,85){\small $\tau_3$}
\put(121,40){\small $\sigma_1$}
\put(225,15){\includegraphics[height=4.0cm,width=0.4\textwidth]
{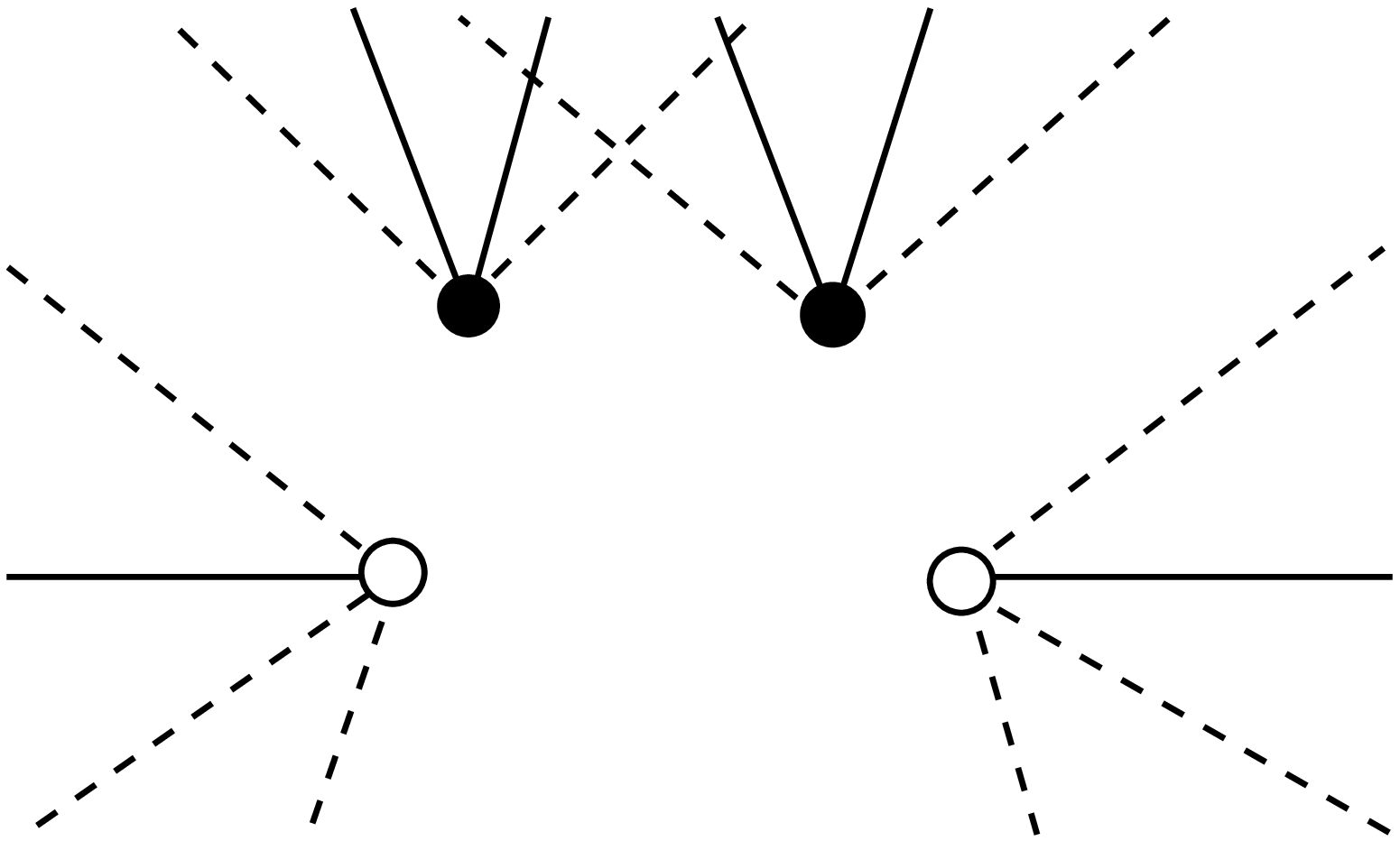}}
\put(293,80){\small $\tau_1$}
\put(338,80){\small $\tau_2$}
\put(278,38){\small $\sigma_1$}
\put(335,38){\small $\sigma_2$}
\end{picture}
\caption{{\em Left}: Cavity graph for $k=3$ with $3$ $\tau$-spins
and $1$ $\sigma$-spin. This graph can be iterated by adding a $\sigma_0$-spin
and connecting it with $3$ long-range bonds (dashed) to the $\tau$-spins and 
$1$ sort-range bond (solid) to the $\sigma$-spin.
{\em Right}: Cavity graph for $k=3$ with $2$ $\tau$-spins and $2$
$\sigma$-spins. This 
graph can be iterated by adding a $\tau_0$-spin and connecting it with $2$
long-range bonds (dashed) to the $\tau$-spins and $2$ short-range bonds
(solid) to the $\sigma$-spins.}
\label{fig:taucav}
\end{figure}

The key difference when  applying the cavity method to the small world
lattice, is that
there are now two different types of cavity spin. Either one can 
remove a short range bond, or one can remove a long range bond. 
A cavity spin with one short range bond and $k$ long range bonds will be
denoted by $\sigma$
and a cavity spin with two short range bonds and $k-1$ long range bonds by
$\tau$. 
The assumption of the cavity method at the level of RS is now that all 
cavity spins of
type $\tau$ have identical individual statistics, as have all cavity
spins of type $\sigma$. The single spin probabilities are parametrized by
\be
p(\tau_j) \sim \rme^{\beta h_j \tau_j} \qquad\qquad
p(\sigma_j) \sim \rme^{\beta x_j \sigma_j}
\ee
and it is assumed that the distributions of the different effective 
cavity fields
$W(h)$ and $\Phi(x)$ are invariant under the appropriate graph iterations.
These graph operations are illustrated in figure
\ref{fig:taucav}. The left figure denotes the set
of cavity spins, which for a small world lattice with $k=3$, can be linked 
onto a new $\sigma$ cavity spin, the right shows spins which can be
linked onto a new $\tau$ cavity spin.
By considering the partition functions of the new cavity spins 
analogously to (\ref{partsum}),
one may again identify the new cavity fields
\begin{eqnarray}
x_0 = u(J_0, x_1) + \sum_{\ell=1}^k u(J_\ell/\bra k\ket, h_\ell)
\nonumber \\
h_0 = u(J_0, x_1) + u(J_0, x_2) + \sum_{\ell=1}^{k-1} u(J_\ell/\bra k\ket, 
h_\ell)
\end{eqnarray}
giving the set of coupled equations
\begin{eqnarray}
\hspace*{-25mm}
\Phi(x) = \int \rmd x^\prime\Phi(x^\prime)
D_k
\delta[x - u(J_0, x^\prime) - \sum_{\ell=1}^k u(J_\ell/\bra k\ket, h_\ell)] \nonumber \\
\hspace*{-25mm} 
W(h) = \int \rmd x_1 \rmd x_2\Phi(x_1) \Phi(x_2) D_{k-1} 
\delta[h-u(J_0,x_1) - u(J_0, x_2) - 
\sum_{\ell=1}^{k-1} u(J_\ell/\bra k\ket, h_\ell)]
\end{eqnarray}
Thus an $x$ cavity field receives one message along the spin chain and
$k$ long range messages, while an $h$ cavity field receives two
messages along the chain and $k-1$ long range messages.

\section{Calculating replica symmetric observables}
To calculate observables with the cavity method, various operations on 
cavity graphs can be defined. In some cases these operations are not unique, 
however, the results they give are equivalent.
\subsection{The total free energy}
We define the $G_{N, q_1,q_2}$
cavity graph to be a graph consisting of $N$ spins, where $q_1$ of
these are cavity spins 
of the $\sigma$-type and $q_2$ are cavity spins of the $\tau$-type.
In particular, consider a $G_{N,4,2k}$ graph. By adding $2$ short 
range bonds and $k$ long range bonds, the latter can be converted into 
a $G_{N,0,0}$ 
graph, whereas, adding $2$ sites (and connecting each of 
them with $2$ short range and
$k$ long range bonds) creates a $G_{N+2,0,0}$ graph.
If adding two short-range and $k$ long range links results in a free energy
shift $\Delta F^{(1)}$ and adding a site with its 2 short range and 
$k$ long range links results in 
the free energy shift $\Delta F^{(2)}$, the total free energy per spin of the 
small world lattice is 
\begin{eqnarray}
\overline{f} = \lim_{N\to \infty} \frac{1}{2} \left[F(G_{N+2,0,0}) - F(G_{N,
    0,0})\right]
\nonumber \\
= \overline{\Delta F^{(2)}} - \frac{1}{2}\overline{\Delta F^{(1)}}
\end{eqnarray}
where $\overline{\cdot\cdot\cdot}$ denotes averages over the quenched disorder.
Expressions for the free energy shifts $\Delta F^{(1)}$ and 
$\Delta F^{(2)}$ are obtained from the difference in the logarithm of
the partition sum before and after the graph operation:
\begin{eqnarray}
\hspace*{-25mm}
-\beta \Delta F^{(1)} =  2\log[\cosh(\beta J_0)] + 
\sum_{\ell=1}^k \log [\cosh(\beta J_\ell/\bra k\ket)]
 \nonumber \\
\hspace*{-5mm} + \log[1+\tanh(\beta u(J_0,x_1))\tanh(\beta x_3)] 
+ \log[1+\tanh(\beta 
u(J_0,x_2))\tanh(\beta x_4)]
 \nonumber \\
+ \sum_{\ell=1}^k \log[1+\tanh(\beta u(J_\ell/\bra k\ket, h_\ell)) \tanh(\beta h_{k+\ell})]
\\
\hspace*{-25mm}
-\beta \Delta F^{(2)} =  
\log \left\{ \frac{\cosh^2(\beta J_0)}{\cosh(\beta u(J_0, x_1))
\cosh(\beta u(J_0, x_2))} \right\}
+ \sum_{\ell=1}^{k}\log \left\{ \frac{\cosh(\beta J_\ell/\bra k\ket)}
{\cosh(\beta u(J_\ell/\bra k\ket, h_\ell))}
\right\} \nonumber \\
+ \log 2\cosh\left[\beta (u(J_0,x_1) + u(J_0,x_2) + \sum_{\ell=1}^k
  u(J_\ell/\bra k\ket, h_\ell)) \right]
\end{eqnarray}
It follows that the total free energy per spin on a graph with long range
fixed connectivity $k$ is given by
\begin{eqnarray}
\hspace*{-10mm}
\beta \overline{f} & = &-\log 2 - \log [\cosh(\beta J_0)] -
\frac{k}{2} \bra \log \cosh(\beta J_\ell/\bra k\ket)\ket_J \nonumber \\
&&+ \int\! \rmd x \rmd x^\prime \Phi(x) \Phi(x^\prime)\log[1+ \tanh(\beta J_0)\tanh(\beta x)
\tanh(\beta x^\prime)]  \nonumber \\
&&+ 2\int\! \rmd x\, \Phi(x) \log \cosh(\beta u(J_0,x)) \nonumber \\
&&+ \frac{k}{2} \int\! \rmd h \rmd h^\prime W(h) W(h^\prime) \bra \log[ 1 + \tanh(\beta J/\bra k\ket)
\tanh(\beta h) \tanh(\beta h^\prime)] \ket_J \nonumber \\
&&+ k \int\! \rmd h\, W(h) \bra \log \cosh
(\beta u(J/\bra k\ket,h)) \ket_J \nonumber \\
&&- \int\! \rmd h \rmd x \rmd x^\prime \log\cosh(\beta h) \Phi(x) 
\Phi(x^\prime)
D_k \nonumber \\
&&\times~ \delta[h-u(J_0, x) - u(J_0,x')
-\sum_{\ell=1}^k u(J_\ell/\bra k\ket,h_\ell)]
\label{eq:free_e_fixed}
\end{eqnarray}
where we have abbreviated $\bra f(J) \ket_J = \int\! \rmd J p_J(J) f(J)$.
We note once more that one can define different sets of graph operations, 
which lead to different formulations of equivalent expressions of
the free energy. 

\subsection{Magnetizations, correlations and graph iterations}
To calculate the magnetization and spin glass order parameters,
the effective field of a normal spin is needed, 
as opposed to a cavity spin. To that end, 
$k$ long range and $2$
short range bonds are linked with a new site, 
which then has the correct set of neighbours.
Its real effective field distribution follows once the
cavity field distributions are known, giving
\begin{eqnarray}
m &= &\int\! \rmd H\, R(H) \tanh(\beta H) \\
q &= & \int\! \rmd H\, R(H)\tanh^2(\beta H) 
\end{eqnarray}
where
\begin{eqnarray}
\hspace*{-20mm}
R(H) &= &
\int\! \rmd x \rmd x^\prime\, \Phi(x) \Phi(x^\prime) D_k ~ 
\delta[H-u(J_0, x) - u(J_0,x^\prime)
-\sum_{\ell=1}^k u(J_\ell/\bra k\ket,h_\ell)]
\end{eqnarray}
To calculate the nearest neighbour correlation function on the ring 
\begin{equation}
C = \lim_{N \rightarrow \infty} \frac{1}{N} \sum_i \overline{\bra \sigma_i \sigma_{i+1} \ket}
\label{nncorf}
\end{equation} 
one can link 
two $\sigma$ cavity spins by a short range bond of strength $J_0$. The
Hamiltonian giving the partition function of the two spins is defined by
\begin{equation}
H(\sigma_1, \sigma_2) = -(J_0 \sigma_1 \sigma_2 + x_1 \sigma_1  + x_2\sigma_2)
\end{equation}
The ensemble averaged nearest neighbour correlation function follows:
\begin{equation}
C = \int\! \rmd x \rmd x^\prime\, \Phi(x)\Phi(x^\prime) 
\frac{\tanh(\beta J_0) + \tanh(\beta x)\tanh(\beta x^\prime)}
{1+ \tanh(\beta J_0) \tanh(\beta x)\tanh(\beta x^\prime)}
\end{equation}
Although giving more cumbersome expressions, other
higher order correlations may be calculated in a similar way.

Anticipating later calculations, we write down expressions for the free
energy shifts resulting from  
adding a $\sigma$ cavity spin to the graph or adding a $\tau$ cavity
spin to the graph. These free energy shifts will be important at the
level of 1RSB:
\begin{eqnarray}
-\beta \Delta F_{\sigma} 
& = & \log \left\{ \frac{\cosh(\beta J_0)}{\cosh(\beta u(J_0, x_1))} \right\}
+ \sum_{\ell=1}^k \log \left\{ \frac{\cosh(\beta J_\ell/\bra k\ket)}
{\cosh(\beta
  u(J_\ell/\bra k\ket, h_\ell))} 
\right\} \nonumber \\
&&+ \log 2\cosh\left[\beta (u(J_0,x_1) + \sum_{\ell=1}^k u(J_\ell/\bra k\ket, h_\ell))\right]
\end{eqnarray}
and
\begin{eqnarray}
\hspace*{-25mm}
-\beta \Delta F_{\tau} = 
\log \left\{ \frac{\cosh^2(\beta J_0)}{\cosh(\beta u(J_0, x_1))
\cosh(\beta u(J_0, x_2))} \right\}
+ \sum_{\ell=1}^{k-1}\log \left\{ \frac{\cosh(\beta J_\ell/\bra k\ket)}
{\cosh(\beta u(J_\ell/\bra k\ket, h_\ell))}
\right\} \nonumber \\
+ \log 2\cosh\left[\beta (u(J_0,x_1) + u(J_0,x_2) + \sum_{\ell=1}^{k-1} 
u(J_\ell/\bra k\ket, h_\ell))
\right]
\end{eqnarray}

\section{`Small world' lattices with fluctuating connectivity}
\subsection{Arbitrary long range connectivity distribution}
We now return to the case of arbitrary connectivity distribution $p(k)$. 
The only modification is that we must now average over the ensemble of
graphs, described by their connectivities $p(k)$, for the different
graph iterations. 
For the iteration of a $\tau$ cavity graph, one should take into account the
degeneracy, i.e., for a given $k$ there are $k$ different ways to remove one
link, and thus each cavity graph with $k-1$ $\tau$ spins is to be weighted
by an additional factor $k$. After normalisation one finds for the 
cavity field distributions
\begin{eqnarray}
\hspace*{-25mm}
\label{eq:cavityx}
\Phi(x) = \int\! \rmd x^\prime\,\Phi(x^\prime)
\sum_{k=0}^\infty p(k)D_k~ 
\delta[x - u(J_0, x^\prime) - \sum_{\ell=1}^k u(J_\ell/\bra k\ket, h_\ell)] \\
\hspace*{-25mm}
W(h) = \int\! \rmd x \rmd x^\prime \,\Phi(x) \Phi(x^\prime) 
\sum_{k=0}^{\infty}
 \frac{p(k)k}{\langle k \rangle} ~D_{k-1}\nonumber \\
\times ~\delta[h-u(J_0,x) - u(J_0, x^\prime) - 
\sum_{\ell=1}^{k-1} u(J_\ell/\bra k\ket, h_\ell)]
\label{eq:Whpk}
\end{eqnarray}
Similarly, the free energy (\ref{eq:free_e_fixed}) can easily be generalised
to the case of a general $p(k)$.

The self-consistent equations for the distributions of the cavity fields admit
the trivial solutions $W(z) = \Phi(z) = \delta(z)$. At high temperature these
are the only solutions and they describe a paramagnetic state. Second
order transitions 
to ferromagnetic or spin glass phases can be found along the lines of
\cite{Ni04}, 
by expansion in small moments around the paramagnetic solution.  
Bifurcations of $\int\! \rmd h\, W(h) h$ and $\int\! \rmd x\, \Phi(x) x$
correspond to a  
transition into the ferromagnetic phase. This will occur at the 
critical temperature solving
\begin{equation}
1=\left[ \langle k \rangle (\rme^{2\beta J_0}-1) + \frac{\langle k^2 \rangle
- \langle k \rangle }{\langle k \rangle} \right] \left\bra 
\tanh\left(\frac{ \beta J}{\bra k\ket} \right) \right\ket_J
\label{eq:fbif}\end{equation}
Similarly, a spin glass phase appears when one of the second moments
$\int\! \rmd h\, W(h) h^2$ or $\int\! \rmd x\, \Phi(x) x^2$
bifurcates, given by the condition
\begin{equation}
1 = \left[ 2\langle k \rangle  \sinh^2(\beta J_0) + \frac{\langle k^2 \rangle 
- \langle k \rangle}{\langle k \rangle} \right] \left\bra \tanh^2\left(
\frac{\beta J}{\langle k \rangle}\right) \right\ket_J 
\label{eq:sgbif}
\end{equation}
\subsection{Correspondence with the replica method}
\label{sec: RSlink}
For models where the number of long range connections
at a given site is Poisson distributed, i.e $p(k) = \rme^{-c}c^k/k!$ one finds that
$\sum_k p(k)k f(k-1)/\langle k\rangle  = \sum_k p(k) f(k)$. We can exploit this
property and recover the results of \cite{Ni04} obtained with the replica approach.
In particular, the replica
symmetric order parameters are the distributions $\tPhi(x),\tPsi(y),\tW(h)$ which
are the solution of the following coupled equations:
\begin{eqnarray}
\label{eq: phi_rs}
  \tPhi(x)&=&\int\!\mathrm{d}x^\prime\,\tPhi(x^\prime)\sum_{k}\frac{\rme^{-c}c^{k}}{k!} 
    D_k~
       \delta[x-\sum_{\ell=1}^ku(J_{\ell}/c,h_{\ell})-u(J_{0},x^\prime)] 
\end{eqnarray}
\begin{eqnarray} 
\label{eq: psi_rs} 
  \tPsi(y)&=&\int\!\mathrm{d}y^\prime\,\tPsi(y^\prime)\sum_{k}\frac{\rme^{-c}c^{k}}{k!} 
    D_k~
       \delta[y-u(J_{0},y^\prime+\sum_{\ell=1}^k u(J_{\ell}/c,h_{\ell}))] 
\end{eqnarray} 
\be 
\label{eq: w_rs} 
  \tW(h)=\int\!\mathrm{d}x\mathrm{d}y\,\tPhi(x)\tPsi(y)\delta[h-x-y] 
\ee
We can immediately see that (\ref{eq: phi_rs}) is exactly the same as the equation of the 
distribution of the $x$-type cavity field (\ref{eq:cavityx}). To verify that $\tW(h)$ 
in (\ref{eq: w_rs}) is the distribution
of the cavity fields of type $h$, we first note 
the following relation
between $\tPhi$ and $\tPsi$:
\be
\label{eq: rs_xfy}
  \tPhi(x)=\sum_{k}\frac{\rme^{-c}c^{k}}{k!}\int\!\mathrm{d}y\,\tPsi(y) 
     D_k ~
       \delta[x-y-\sum_{\ell=1}^{k}u(J_{\ell}/c,h_{\ell})]
\ee
which can be easily verified by inserting this expression for $\tPhi$ in
(\ref{eq: phi_rs}) and then using (\ref{eq: psi_rs}).
Upon substituting $\tPhi,\tPsi$ in the right-hand side of (\ref{eq: w_rs}) with
(\ref{eq: phi_rs}), (\ref{eq: psi_rs}) and introducing the shorthand
$p_{k}=\rme^{-c}c^{k}/k!$ we get:
\begin{eqnarray}
  \tW(h)
   &=&\int\!\mathrm{d}x^\prime\mathrm{d}x^{\prime \prime}\,\tPhi(x^\prime)
      \bigg\{\sum_{k^\prime}\int\!\mathrm{d}y^\prime\,\tPsi(y^\prime)D_{k^\prime}~
\delta[x^{\prime\prime}-y^\prime-\sum_{\ell^\prime=1}^ku(J_{\ell^\prime}/c,h_{\ell^\prime})]
      \bigg\} \nonumber \\
   &&\times
     \sum_{k}p_{k}\int\! D_{k}\,\delta[h-u(J_{0},x^\prime)-u(J_{0},x^{\prime\prime})-
\sum_{\ell=1}^ku(J_{\ell}/c,h_{\ell})]
\end{eqnarray}
From (\ref{eq: rs_xfy}) it follows that the expression in the curly
brackets above is $\tPhi(x^{\prime\prime})$, therefore 
we get the same equation for the field $h$ as from the cavity approach 
(given the Poisson distribution of the long range connectivity).
Thus we conclude that the cavity field distribution $W(h)$ is identical
to the effective field distribution $\tW(h)$ 
(defined via $W(h)=\lim_{N\to\infty}\frac{1}{N} \sum_i \overline{\delta[h-
\tanh^{-1}(\bra\sigma_{i}\ket)]}$) on the Poisson random graph.
This correspondence between the two methods also provides a nice physical
interpretation of the $n\to 0$ limit of the eigenvalue problem of the replicated
transfer matrix of the replica approach described by equation
(\ref{eq: phi_rs}). The right  $n\to 0$ eigenvector,
associated with the largest eigenvalue is the type $x$ cavity field distribution. 

Finally let us comment that
when the free energy (\ref{eq:free_e_fixed}) is generalized
to the case of a Poissonian connectivity, 
the self-consistency equation (\ref{eq:Whpk}) for $W(h)$ can be inserted 
into the last term to give
\small
\begin{eqnarray}
\hspace*{-20mm}
\beta f = - \log 2 - \log [\cosh(\beta J_0)] -
\frac{c}{2} \bra \log \cosh(\beta J/\bra k\ket) \ket_J \nonumber \\
\hspace*{-10mm}
+ \int\! \rmd x \rmd x^\prime \, \Phi(x) \Phi(x^\prime)\,\log[1+ \tanh(\beta J_0)\tanh(\beta x)
\tanh(\beta x^\prime)]  \nonumber \\
\hspace*{-10mm}- \int\! \rmd x\, \Phi(x) \log
[1-\tanh^2(\beta J_0) \tanh^2(\beta x)] \nonumber \\
\hspace*{-10mm}
+ \frac{c}{2} \int\! \rmd h \rmd h^\prime \, W(h) W(h^\prime) \bra 
\log[ 1 + \tanh(\beta J/\bra k\ket)
\tanh(\beta h) \tanh(\beta h^\prime)] \ket_J \nonumber \\
\hspace*{-10mm}
- \frac{c}{2} \int\! \rmd h W(h) \bra \log [1-\tanh^2(\beta J_\ell/\bra k\ket)\tanh^2(\beta h)]
\ket_J - \int\! \rmd h W(h) \log \cosh(\beta h)
\label{fe}
\end{eqnarray}
\normalsize
It is then straightforward to verify that this expression is equivalent to the result
in \cite{Ni04}.

\section{1RSB cavity solution of the `small-world' spin glass}
Considering that the model we have described so far is still
general in allowing an arbitrary site-independent 
connectivity distribution $p(k)$ and distribution of 
interaction strengths $p_J(J)$ \footnote{Of course the distributions should
be sufficiently well-behaved to give a well-defined free energy} , 
it is beyond doubt that cases in which replica symmetry is broken are 
within the possible parameter regimes.
An example is the case where $p(k)=\delta_{k,6}$, 
$p_J(J) = \frac{1}{2}\delta(J-6) + \frac{1}{2}\delta(J+6)$ and $J_0 = 0$, 
exactly the Bethe
lattice spin glass studied in \cite{MePa01} (note that we have
rescaled the fields in our model definitions).
Having formulated the replica symmetric solution of the model on the small
world lattice using the cavity method, we proceed by applying the formalism 
explained carefully in
\cite{MePa01}, in order to derive results at the level of 1RSB. Since
the generalization of the 1RSB cavity  
method for the Bethe lattice to the small world lattice is straightforward 
once the RS results are known, we will just briefly give the resulting 
1RSB equations for the small world lattice and discuss the differences
with the Bethe lattice.  

\subsection{Basic assumptions of the 1RSB formulation}
Instead of assuming that the state with the lowest free energy, 
characterised by $W(h)$ and $\Phi(x)$, is
invariant under cavity graph iterations of any kind, at the level of
1RSB one considers the $\cM$ locally stable states  
of lowest free energy. At a cavity site $i$, one defines the effective 
cavity fields $h^\alpha_i$ or $x^\alpha_i$, where 
$\alpha \in \{1, \ldots, \cM\}$ labels the particular state. To each 
state $\alpha$ a free energy $F^\alpha$ is associated. The crucial point is
that upon a graph iteration each pure state $\alpha$ 
receives a different free energy shift $\Delta F^\alpha$, which is a 
function of the $\{h^\alpha_i\}$ and $\{x^\alpha_j\}$ 
being linked to the new site. The ordering of new free energies of the 
different states
after a graph iteration need not be the same as that before.
The distribution of free energies is postulated to be of the form

\begin{equation}
\rho(F) = \exp(\beta \mu (F - F^{ref}))
\label{fedist}
\end{equation}
and the likelihood $W^\alpha$ for the system of being in a state $\alpha$ 
is given by
\begin{equation}
W^\alpha = \frac{\exp(-\beta F^\alpha)}{\sum_\gamma \exp(-\beta
  F^\gamma)}
\end{equation}
The distributions of the different cavity field distributions are 
assumed to factorize over pure states at each site in the following
manner:
\begin{equation}
P(\bh) = \frac{1}{N} \sum_i \prod_{\alpha = 1}^\cM P_i(h^\alpha)
\label{eq:Phrsb}
\end{equation}
\begin{equation}
Q(\bx) = \frac{1}{N} \sum_i \prod_{\alpha = 1}^\cM Q_i(x^\alpha)
\label{eq:Qxrsb}
\end{equation}
Similarly, one can define the distribution of effective local fields on
a normal (non-cavity) spin:
\begin{equation}
O(\bH) = \frac{1}{N} \sum_i \prod_{\alpha = 1}^\cM O_i(H^\alpha)
\label{eq:OHrsb}
\end{equation}
Then the cavity and effective field distributions for a pure state are 
reweighted by the free energy shifts upon a graph iteration:
\begin{eqnarray}
P_0(h_0) = C^{-1}_{P} \int\! \rmd (\Delta F_\tau)\, W(h_0, \Delta F_\tau) \exp(-\beta
\mu \Delta F_\tau)
\label{eq:Pdist} \\
Q_0(x_0) = C^{-1}_{Q} \int\! \rmd (\Delta F_\sigma)\, \Phi(x_0, \Delta F_\sigma) \exp(-\beta
\mu \Delta F_\sigma)
\label{eq:Qdist} \\
O_0(H_0) = C^{-1}_{O} \int\! \rmd (\Delta F^{(2)})\, R(H_0, \Delta F^{(2)}) \exp(-\beta \mu
\Delta F^{(2)})
\end{eqnarray}
Here $W(h_0, \Delta F_\tau)$, $\Phi(x_0, \Delta F_\sigma)$ and 
$R(H_0, \Delta F^{(2)})$ denote the joint distributions of fields and 
free energy shifts. Since one has
\begin{eqnarray}
h_0^\alpha = h_0^\alpha(h_1^\alpha,\ldots,h_{k-1}^\alpha,x_1^\alpha,x_2^\alpha,
J_1, \ldots, J_{k-1})
\nonumber \\
\Delta F_\tau^\alpha=\Delta F_\tau^\alpha(
h_1^\alpha,\ldots,h_{k-1}^\alpha,x_1^\alpha,x_2^\alpha,
J_1, \ldots, J_{k-1}) \nonumber \\
x_0^\alpha = x_0^\alpha(h_1^\alpha,\ldots,h_{k}^\alpha,x_1^\alpha,
J_1, \ldots, J_{k})
\nonumber \\
\Delta F_\sigma^\alpha=\Delta F_\sigma^\alpha(
h_1^\alpha,\ldots,h_{k}^\alpha,x_1^\alpha,
J_1, \ldots, J_{k}) \nonumber \\
H_0^\alpha = H_0^\alpha(h_1^\alpha,\ldots,h_{k}^\alpha,x_1^\alpha,x_2^\alpha,
J_1, \ldots, J_{k})
\nonumber \\
\Delta F^{(2)\alpha}=\Delta F^{(2)\alpha}(
h_1^\alpha,\ldots,h_{k}^\alpha,x_1^\alpha,x_2^\alpha,
J_1, \ldots, J_{k}) 
\end{eqnarray}
these joint distributions are obtained by linking the appropriate 
sets of 
cavity spins in the same pure state $\alpha$ onto a new spin.

Alternatively, one could formulate the iteration of distributions in terms of 
iterations of functionals $\mathcal{F}[\{Q\}]$ and
$\mathcal{W}[\{P\}]$, representing functional densities of the 
distributions (\ref{eq:Pdist}) and (\ref{eq:Qdist}), giving
\small
\begin{eqnarray}
\label{eq: functionalQ}
\hspace*{-25mm}
\mathcal{F}[\{Q\}]=\sum_k p_k \int\! \{\rmd Q'\}\mathcal{F}[\{Q'\}] \prod_{\ell \leq k} 
\left[\int\! \{\rmd P_\ell\}
  \rmd J_\ell \mathcal{W}[\{P_\ell\}] p_{J}(J_\ell)\right]\nonumber\\
\prod_x \delta\left\{Q(x) - \frac{1}{C_{Q}} \int\! \rmd x'\, Q'(x')
\prod_{\ell \leq k}
  [\int\!\rmd h_\ell P_{\ell}(h_\ell)]\, \right. \nonumber \\
 \hspace*{45mm} \left. \times \delta\left[x - u(J_0, x') - \sum_{\ell\leq k}
  u(J_\ell/\bra k \ket,h_\ell)\right]\rme^{\beta \mu \Delta F_\sigma}\right\}
\end{eqnarray}
\normalsize

\begin{eqnarray}
\label{eq: functionalP}
\hspace*{-25mm}
\mathcal{W}[\{P\}]=
\sum_k \frac{p_k k}{\bra k\ket}\int\! \{\rmd Q\}\{\rmd Q'\}\,\mathcal{F}[\{Q\}]
\mathcal{F}[\{Q'\}] 
 \prod_{\ell \leq k-1} \left[\int\! \{\rmd P_\ell\}
  \rmd J_\ell \mathcal{W}[\{P_\ell\}] p_{J}(J_\ell)\right] 
\nonumber \\ 
\prod_h \delta\left\{P(h) - \frac{1}{C_{P}} \int\! \rmd x \rmd x'\, Q(x)
  Q'(x') \prod_{\ell\leq k-1} [\int\!\rmd h_\ell P_{\ell}(h_\ell)] \right. \nonumber\\
\hspace*{15mm}\left. \times~ \delta\left[h - u(J_0, x)
  -u(J_0, x') -  \sum_{\ell\leq k-1}
  u(J_\ell/\bra k\ket,h_\ell)\right] \rme^{\beta \mu \Delta F_\tau}\right\}
\end{eqnarray}
where $C_{Q},C_{P}$ are normalization constants.

\subsection{Observables in 1RSB formulation}
Numerically solving the 1RSB equations can be achieved with
a population dynamics algorithm. The distributions 
(\ref{eq:Phrsb},\ref{eq:Qxrsb},\ref{eq:OHrsb}) 
are approximated by $\cN$ populations
of $\cM$ fields. The result of a sufficient number of iterations of 
these $\cN$ populations is a stochastically stable set of populations of fields.
The various observables are found from these populations of 
fields via the following equations:
\begin{equation}
m = \frac{1}{\cM} \sum_\alpha \tanh(\beta H^\alpha)
\end{equation}
\begin{equation}
q_1 = \frac{1}{\cM} \sum_\alpha \tanh^2(\beta H^\alpha)
\end{equation}
\begin{equation}
q_0 =\frac{1}{\cM(\cM -1 )} \sum_{\alpha \neq \beta} \tanh(\beta
H^\alpha) \tanh(\beta H^\beta)
\end{equation}
\begin{eqnarray}
\hspace*{-5mm} f = 
\frac{1}{2\beta \mu}  \ln \left[\frac{1}{\cM} \sum_{\alpha}
\exp(-\beta \mu \Delta F_\alpha^{(1)}) \right] - \frac{1}{\beta \mu} \ln \left[ \frac{1}{\cM} \sum_{\alpha}
\exp(-\beta \mu \Delta F_\alpha^{(2)})\right] 
\end{eqnarray}
Note that in the above expressions, $\Delta F_\alpha^{(1)}$ and
$\Delta F_\alpha^{(2)}$ are functions of the fields $\{ h^\alpha_i\}$ 
and $\{x^\alpha_i\}$, as well as $\{ J_i\}$. The final results are 
understood to be averages of many samples from the populations of fields
and distributions $p_{J}(J)$.
The parameter $\mu$ corresponds to the fraction of overlaps equal to
$q_0$ and should thus have a value between $0$ and $1$. The correct value is
found by extremization of the free energy. Thus the right value of $\mu$
is the one for which the derivative of the free energy with 
respect to $\mu$ is zero. The latter is expressed as
\begin{equation}
\frac{\partial f}{\partial \mu} = d^{(2)}- \frac{1}{2} d^{(1)}
\end{equation}
where
\begin{equation}
d^{(1)}\equiv\frac{\partial F^{(1)}}{\partial \mu}= \frac{-1}{\mu} F^{(1)} 
+\frac{1}{\mu}
\frac{\sum_\alpha \exp(-\beta \mu \Delta F_\alpha^{(1)}) \Delta
  F_\alpha^{(1)} }{\sum_\alpha \exp(-\beta \mu \Delta F_\alpha^{(1)})}
\end{equation}
and 
\begin{equation}
d^{(2)}\equiv\frac{\partial F^{(2)}}{\partial \mu}
= \frac{-1}{\mu} F^{(2)} +\frac{1}{\mu}
 \frac{\sum_\alpha \exp(-\beta \mu \Delta F_\alpha^{(2)}) \Delta
  F_\alpha^{(2)} }{\sum_\alpha \exp(-\beta \mu \Delta F_\alpha^{(2)})}
\end{equation}
By defining the free energy shift of a short range linking of two 
$\sigma$ cavity spins
\begin{equation}
-\beta \Delta F_b^\alpha = \log \sum_{\sigma_1,\sigma_2} \exp
 \left\{-\beta 
 H(\sigma_1,\sigma_2;x_1^\alpha, x_2^\alpha) \right\}
\end{equation}
one may write the expression for the nearest neighbour 
correlation function as
\begin{equation}
C = \frac{\sum_\alpha \exp[-\beta \mu \Delta F_b^\alpha]\frac{\tanh(\beta
  J_0) + \tanh(\beta x_1^\alpha) \tanh(\beta x_2^\alpha)}
{1 + \tanh(\beta J_0)
  \tanh(\beta x_1^\alpha) \tanh(\beta x_2^\alpha)}}
{\sum_\alpha \exp[-\beta \mu \Delta
  F_b^\alpha] }
\end{equation}

\section{1RSB replica solution of the `small-world' spin glass}
\subsection{General remarks}
In this section we consider the spin glass model on a `small-world' lattice with a Poisson
distribution for the number of long range connections at a given site. This model was
studied  with the replica formalism in \cite{Ni04}, where phase diagrams where
derived using the replica symmetric ansatz. The basic ingredients of
this approach are the the $2^{n}\times 2^{n}$ replicated transfer
matrix $\bT[P]$ with entries:
\be
\label{eq: rtm}
  T_{\bsigma\bsigma^\prime}[P]=\rme^{\beta J_{0}\bsigma\bsigma^\prime+
      c\sum_{\btau}P(\btau)\bra \rme^{\frac{\beta J}{c}\btau\bsigma}\ket_{J}-c}
\ee
and the order parameter function $P(\btau)$, which in the limit $N\to\infty$
gives the fraction of the $n$-replicated spins
$\bsigma_{i}=(\sigma_{i}^{1},\ldots,\sigma_{i}^{n})$ with a given configuration
$\btau\in\{-1,1\}^{n}$, and satisfies the self-consistent equation:
\be
\label{eq: Psigma}
  P(\bsigma)=\frac{v_{\bsigma}[P]u_{\bsigma}[P]}
                  {\sum_{\bsigma^\prime}v_{\bsigma^\prime}[P]u_{\bsigma^\prime}[P]}
\ee
where $\bv[P],\bu[P]$ are the left and right eigenvectors associated with the
largest eigenvalue of the replicated transfer matrix $\bT[P]$.

\subsection{Eigenvalue problem for the 1RSB replicated transfer matrix}
We consider the first step of Parisi's replica symmetry breaking
scheme \cite{MePa87}, according to which the $n$ different replicas
are divided into $n/m$
blocks of $m$ replicas and 
in the limit $n\to 0$ the integer $m$ becomes a real number between 0
and 1. 
In finite connectivity models this results in the order parameter function
$P(\bsigma)$ acquiring the form \cite{monasson98}:
\be
\label{eq: 1rsb_Psigma}
  P(\bsigma)=\int\!\{\rmd P\}\,\mathcal{W}[\{P\}]\prod_{l=1}^{n/m}
    \int\!\mathrm{d}h\,P(h)\frac{\rme^{\beta h\sum_{a=1}^{m}\sigma_{a,l}}}
                              {[2\cosh(\beta h)]^{m}}
\ee
The physical interpretation of the functional $\mathcal{W}[\{P\}]$ is similar to
that of the replica symmetric function $W(h)$. 
In particular one considers effective fields defined via $h_{i}^{\alpha}=\frac{1}{\beta}
\arctanh\bra\sigma_{i}\ket_{\alpha}$, but in contrast with the RS case the existence of
different pure states or ergodic components (denoted by $\alpha$) is now taken into
account. The effective field, at a given site $i$, fluctuates due to
the presence of different pure states. If the probability distribution
    of this field is $P_{i}(h)$ 
then this distribution will generally be different from site to site, therefore the relevant
order parameter is the functional probability measure $\mathcal{W}[\{P\}]$ of those 
densities.
Insertion of (\ref{eq: 1rsb_Psigma}) in the general expression for the replicated transfer
matrix (\ref{eq: rtm}) leads, after some straightforward manipulations, to the following
expression for the 1RSB replicated transfer matrix:
\be
\label{eq: rtm_1rsb}
  T_{\bsigma\bsigma^\prime}^{1RSB}=
    \int\!\{\rmd M\}\,\mathcal{P}[\{M\}]\prod_{l=1}^{n/m}
      \int\!\mathrm{d}\theta\,M(\theta\vert m)
      \rme^{\beta J_{0}\sum_{a=1}^{m}\sigma_{a,l}\sigma_{a,l}^\prime+
         \beta\theta\sum_{a=1}^{m}\sigma_{a,l}}
\ee
where the shorthand $\mathcal{P}$ which will be used in the remainder of this section
is defined as:
\begin{eqnarray}
\hspace*{-25mm}
\label{eq: Qq}
  \mathcal{P}[\{M\}]=\sum_{k}\frac{\rme^{-c}c^{k}}{k!}
  \prod_{\ell=1}^{k}
   \left\{ \int\!\mathrm{d}J_{\ell}\,p_{J}(J_{\ell}) \{\rmd P_\ell\}\,\mathcal{W}[\{P_\ell\}] \right\}
  \nonumber \\
  \hspace*{-20mm}\times
    \prod_{\theta}\delta\left\{
      M(\theta\vert m)-
        \prod_{\ell=1}^{k}\left[\int\!\frac{\mathrm{d}h_\ell\,P_\ell(h_\ell)}
                                                 {[2\cosh(\beta h_\ell)]^{m}}\right]
      \rme^{\beta m\sum_\ell B(J_\ell/c,h_\ell)}\,
      \delta[\theta-\sum_{\ell=1}^{k}u(J_\ell/c,h_\ell)]\right\}
\end{eqnarray}
with
\bd 
  B(J,z)=\frac{1}{2\beta}\log\left\{ 4\cosh[\beta(J+z)]\cosh[\beta(J-z)]\right\}
\ed
We observe that with each of the $n/m$ blocks (labeled by $l$) a
different $2^{m}\times 2^{m}$ matrix 
is associated. This may be viewed as a transfer matrix of a one dimensional chain
with random fields $\theta$. Note however that $M(\theta\vert m)$ is not normalized
(as would be the case in a 
purely $1D$ model). The complete matrix consists of the $n/m$-fold Kronecker product of the
$2^{m}\times 2^{m}$ matrices averaged over the functional measure
$\mathcal{P}$, which gives the 
probability of drawing a particular `distribution' of random fields $M(\theta\vert m)$.

We proceed as in the replica symmetric approach and introduce an
ansatz for the left and right eigenvectors associated with the
largest eigenvalue of the replicated transfer matrix (\ref{eq: rtm_1rsb})
which is the natural generalization of the replica symmetric eigenvectors
(see e.g \cite{nik-coolen04},\cite{Ni04}) 
\be
\label{eq: right_vec}
  u^{1RSB}_{\bsigma}=\int\!\{d\phi\}\,U[\{\phi\}\vert n]
   \prod_{l=1}^{n/m}\int\!\mathrm{d}x\,\phi(x)
    \frac{\rme^{\beta x\sum_{a=1}^{m}\sigma_{a,l}}}{[2\cosh(\beta x)]^{m}}
\ee
\be
\label{eq: left_vec}
  v^{1RSB}_{\bsigma}=\int\!\{d\psi\}\,V[\{\psi\}\vert n]
   \prod_{l=1}^{n/m}\int\!\mathrm{d}y\,\psi(y)
   \frac{\rme^{\beta y\sum_{a=1}^{m}\sigma_{a,l}}}{[2\cosh(\beta y)]^{m}}
\ee
where $\phi(x)\,,\psi(y)$ are assumed to be normalized.
Insertion in the eigenvalue equations (to be satisfied for every $\bsigma$):
\bd
\label{eq: eigen}
\hspace*{-15mm}
  \sum_{\bsigma^\prime}T_{\bsigma\bsigma^\prime}^{1RSB}u_{\bsigma^\prime}^{1RSB}=
    \lambda_{1RSB}(n)u_{\bsigma}^{1RSB}
  \qquad
  \sum_{\bsigma^\prime}v_{\bsigma^\prime}^{1RSB}T_{\bsigma^\prime\bsigma}^{1RSB}=
    \lambda_{1RSB}(n)v_{\bsigma}^{1RSB}
\ed
leads to  new eigenvalue problems for the functionals $U[\{\phi\}\vert n],V[\{\psi\}\vert n]$:
\be
\label{eq: right}
  \lambda_{1RSB}(n)U[\{\phi\}\vert
    n]=\int\!\{\rmd \phi^\prime\}\,\Lambda_{U}(\phi,\phi^\prime\vert n)U[\{ \phi^\prime \}\vert n]
\ee
\be
\label{eq: left}
  \lambda_{1RSB}(n)V[\{\psi\}\vert
    n]=\int\!\{\rmd\psi^\prime\}\,\Lambda_{V}(\psi,\psi^\prime\vert
  n)V[\{\psi^\prime\}\vert n] 
\ee
where
\be
\label{eq: ker_right}
\hspace*{-25mm}
  \Lambda_{U}(\phi,\phi^\prime\vert n)=
    \int\!\{ \rmd
  M\}\,\mathcal{P}[\{M\}]\prod_{x}\,\delta[\phi(x)-\mathcal{A}_{U}(x,\phi^\prime\vert
  M)] 
    \Big[\lambda(m\vert\phi^\prime, M)\Big]^{n/m}
\ee
\be
\hspace*{-25mm}
  \mathcal{A}_{U}(x,\phi^\prime\vert M)=\frac{
    \int\!\mathrm{d}x^\prime\mathrm{d}\theta\,\phi^\prime(x^\prime)M(\theta\vert m)
     \frac{\rme^{m\beta B(J_{0},x^\prime)}}{[2\cosh(\beta x^\prime)]^{m}}
      [2\cosh(\beta x)]^{m}\,\delta[x-\theta-u(J_{0},x^\prime)]}
  {\lambda(m\vert\phi^\prime,M)}
\ee
\be
\label{eq: ker_left}
\hspace*{-25mm}
  \Lambda_{V}(\psi,\psi^\prime\vert n)=
    \int\!\{\rmd M\}\,\mathcal{P}[\{M\}]\prod_{y}\,\delta[\psi(y)-\mathcal{A}_{V}(y,\psi^\prime\vert M)]
    \Big[\lambda(m\vert\psi^\prime, M)\Big]^{n/m}
\ee
\be
\hspace*{-25mm}
  \mathcal{A}_{V}(y,\psi^\prime\vert M)=\frac{
    \int\!\mathrm{d}y^\prime\mathrm{d}\theta\,\psi^\prime(y^\prime)M(\theta\vert m)
    \frac{\rme^{m\beta B(J_{0},y^\prime+\theta)}}{[2\cosh(\beta y^\prime)]^{m}}
     [2\cosh(\beta y)]^{m}\delta[y-u(J_{0},y^\prime+\theta)]}
    {\lambda(m\vert\psi^\prime,M)}
\ee
and
\be
\hspace*{-25mm}
  \lambda(m\vert\psi^\prime,M)= \int \rmd y  
\mathcal{A}_{V}(y,\psi^\prime\vert M)
\hspace*{15mm}
\label{eq: inter_eigval}
  \lambda(m\vert\phi^\prime,M)=\int \rmd x  
\mathcal{A}_{U}(x,\phi^\prime\vert M)
\ee

\subsection{Self-consistent equation for the 1RSB order parameter}
To find a self-consistent equation for the 1RSB order parameter
(\ref{eq: 1rsb_Psigma}) we only need to inspect the $n\to 0$ limit of the
eigenvalue problems (\ref{eq: right},\ref{eq: left}) which are coupled via the
$W[\{P\}]$ dependence of $\mathcal{P}[\{M\}]$. For the free energy however, we
also need to know the $\order(n)$ term of the eigenvalue $\lambda_{1RSB}(n)$.
As in the RS case \cite{Ni04}, we can see by setting $n=0$ in
(\ref{eq: right},\ref{eq: left},\ref{eq: ker_right},\ref{eq: ker_left}),
integrating over $\phi,\psi$ and using $\int \{\rmd \phi\}\,U[\{\phi\}\vert 0]=
\int\{ \rmd \psi\}\,V[\{\psi\}\vert 0]=1$ (since $U[\{\phi\}\vert
  0],V[\{\psi\}\vert 0]$ represent probability 
measures) that: $\lambda_{1RSB}(0)=1$. This in turn implies that the $n=0$ functionals
(which from now on we denote simply by $U[\{\phi\}],V[\{\psi\}]$) are given by:
\be
\label{eq: U}
  U[\{\phi\}]=\int\!\{\rmd \phi^\prime\}\,U[\{\phi^\prime\}]\int\!\{\rmd M\}\,\mathcal{P}[\{M\}]
   \prod_{x}\,\delta\Big[\phi(x)-\mathcal{A}_{U}(x,\phi^\prime\vert M)\Big]
\ee
\be
\label{eq: V}
  V[\{\psi\}]=\int\!\{\rmd \psi^\prime\}\,V[\{\psi^\prime\}]\int\!\{\rmd M\}\,\mathcal{P}[\{M\}]
   \prod_{y}\,\delta\Big[\psi(y)-\mathcal{A}_{V}(y,\psi^\prime\vert M)\Big]
\ee
These equations may be viewed as population dynamics equations for populations
of distributions $\phi(x),\psi(y)$, which are distributed according to
   $U[\{\phi\}],V[\{\psi\}]$, 
with functional update rules $\mathcal{A}_{U},\mathcal{A}_{V}$.
Note that the defining properties of a probability distribution, viz. non-negativity
and normalization are preserved by the update rules as they should be. 

Furthermore, in the limit $n\to 0$ insertion of (\ref{eq:
  1rsb_Psigma},\ref{eq: right_vec},\ref{eq: left_vec}) in (\ref{eq: Psigma}) results in:
\begin{eqnarray}
\label{eq: funcW}
\hspace*{-25mm}
  \mathcal{W}[\{P\}]=\int\!\{\rmd \phi\}\{\rmd \psi\}\,U[\{\phi\}]V[\{\psi\}]
  \nonumber \\
\hspace*{-5mm}\times
    \prod_{h}\,\delta\bigg[P(h)-\frac{[2\cosh(\beta h)]^{m}}{C_{h}(\phi,\psi)}
     \int\!\frac{\mathrm{d}x\mathrm{d}y\,
     \phi(x)\psi(y)}{[4\cosh(\beta x)\cosh(\beta y)]^{m}}\,\delta(h-x-y)\bigg]
\end{eqnarray}
with the normalization constant
\be
  C_{h}(\phi,\psi)=\int\!\mathrm{d}x\mathrm{d}y\,\phi(x)\psi(y)\left[
        \frac{\cosh(\beta x+\beta y)}{2\cosh(\beta x)\cosh(\beta y)}\right]^{m}
\ee
The final triplet of coupled equations (\ref{eq: U},\ref{eq: V},\ref{eq: funcW})
determines the functionals $U,V,W$ which play the role of 1RSB order parameters.

Scalar observables such as magnetization, spin glass order parameters etc, can be
expressed as averages over $P(\bsigma)$:
\be
\label{eq: obs}
  m=\sum_{\bsigma}P(\bsigma)\sigma_{a}\qquad
  q_{ab}=\sum_{\bsigma}P(\bsigma)\sigma_{a}\sigma_{b}
\ee
Note that (\ref{eq: 1rsb_Psigma}) encodes, at the 1RSB level, the hierarchical structure of
the set of pure states which is inherent in the Parisi solution of the infinite range
spin glass. Replicas within the same block $l \in\{1,\ldots,n/m\}$ are assumed to
be in the same pure state, therefore the inter-state overlap 
$q_{1}=\lim_{N\to\infty}\frac{1}{N}\sum_{i}\overline{\bra\sigma_{i}\ket_{\alpha}^{2}}$
is found from (\ref{eq: obs}) with spins $\sigma_{a},\sigma_{b}$ belonging
to the same block $l$:
\be
\label{eq: q1}
  q_{1}=\int\!\{\rmd P\}\,\mathcal{W}[\{P\}]\int\!\mathrm{d}h\,P(h)\tanh^{2}(\beta h)
\ee
All different pure states belong to the same cluster, and the overlap between any two
different pure states defined as
$q_{\alpha\gamma}=\lim_{N\to\infty}\frac{1}{N}\sum_{i}
 \overline{\bra\sigma_{i}\ket_{\alpha}\bra\sigma_{i}\ket_{\gamma}}$, is the same.
This is obtained from (\ref{eq: obs}) with spins $\sigma_{a},\sigma_{b}$ belonging
to different blocks $l\neq l^\prime$:
\be
\label{eq: q0}
  q_{0}=\int\!\{\rmd P\}\,\mathcal{W}[\{P\}]\left[
   \int\!\mathrm{d}h\,P(h)\tanh(\beta h)\right]^{2}
\ee

\subsection{Derivation of the 1RSB free energy}
The general expression of the disordere-averaged free energy per spin in
terms of the order parameter function $P(\bsigma)$ is:
\be
  \bar{f}=\lim_{n\to 0}\frac{1}{\beta n}\bigg\{
   \frac{c}{2}\sum_{\bsigma,\bsigma^\prime}P(\bsigma)P(\bsigma^\prime)
     \bra \rme^{\frac{\beta J}{c}\bsigma\bsigma^\prime}-1\ket_{J}-\log\lambda(n;P)\bigg\}
\ee
where $\lambda(n;P)$ is the largest eigenvalue of the replicated
transfer matrix (\ref{eq: rtm}). 
To find the first contribution in the 1RSB solution, we insert (\ref{eq: 1rsb_Psigma})
into the above and work out the double replicated spin summation. In the limit $n\to 0$ we get:
\begin{eqnarray}
\label{eq: energetic}
\hspace*{-25mm}
  \lim_{n\to 0}\frac{c}{2\beta n}\sum_{\bsigma,\bsigma^\prime}P(\bsigma)P(\bsigma^\prime)
    \Big\bra \rme^{\frac{\beta J}{c}\bsigma\bsigma^\prime}-1\Big\ket_{J}=
    \frac{c}{2\beta m}\int\!\{\rmd P \} \{ \rmd P^\prime\}\,
    \mathcal{W}[\{P\}]\mathcal{W}[\{P^\prime\}]
  \nonumber \\
\hspace*{-20mm}\times
   \left\bra \log\bigg(\int\!\mathrm{d}h\mathrm{d}h^\prime\,P(h)P^\prime(h^\prime)
   \cosh^{m}(\frac{\beta J}{c})[1+\tanh(\frac{\beta J}{c})\tanh(\beta h)\tanh(\beta h^\prime)]^{m}  \bigg)\right\ket_{J}
\end{eqnarray}
The second contribution to the free energy involves the largest
eigenvalue of the 1RSB replicated transfer matrix 
(\ref{eq: rtm_1rsb}) . We have already seen that
$\lambda_{1RSB}(0)=1$, therefore for small $n$ we have:
$\lambda_{1RSB}(n)=1+n\lambda+\order(n^{2})$, which in turn implies that the
contribution of the eigenvalue to the free energy in the limit $n\to 0$ is
$-\frac{\lambda}{\beta}$.  To find $\lambda$ note that from
(\ref{eq: right},\ref{eq: ker_right})  follows:
\be
  \lambda_{1RSB}(n)=
  \frac{\int\!\{\rmd \phi\}\,U[\{\phi\}\vert n]\int\!\{\rmd M\}\,\mathcal{P}[\{M\}]
        \rme^{\frac{n}{m}\log\lambda(m\vert \phi,M)}}
       {\int\!\{\rmd \phi\}\,U[\{\phi\}\vert n]}
\ee
We then expand the right-hand side around $n=0$ to find the $\order(n)$ term.
After some straightforward manipulations and using (\ref{eq: Qq},\ref{eq: inter_eigval})
the second contribution to the 1RSB free energy becomes:
\begin{eqnarray}
\label{eq: entropic}
\hspace*{-25mm}
  \frac{\lambda}{\beta}=\frac{1}{\beta}\log 2\cosh(\beta J_{0})+
   \frac{1}{\beta m}\int\!\{\rmd \phi\}\,U[\{\phi\}]\sum_{k}\frac{\rme^{-c}c^{k}}{k!}
   \prod_{\ell=1}^{k}\left[\int\!\mathrm{d}J_\ell\{\rmd P_\ell\} P(J_\ell)
    \mathcal{W}[\{P_\ell\}]\right]
  \nonumber \\
\hspace*{-20mm}\times
  \log\left\{\int\!\mathrm{d}x\,\phi(x)\prod_{\ell=1}^{k}
\left[ \int\!   \mathrm{d}h_\ell\,P_\ell(h_\ell)\cosh^{m}(\frac{\beta J_\ell}{c}) \right]
\right.
\nonumber\\
\left. 
\left(\frac{1}{2}
\prod_{\ell=1}^k [1+\tanh(\frac{\beta J_\ell}{c})\tanh(\beta h_\ell)]
+ \frac{1}{2}
\prod_{\ell=1}^k [1-\tanh(\frac{\beta J_\ell}{c})\tanh(\beta h_\ell)]
\right)^m  
 \right. 
  \nonumber \\ \left. 
\hspace*{25mm}\times
  \Big[1+\tanh(\beta x)\tanh(\beta J_{0})
       \tanh[\beta\sum_\ell u(J_\ell/c,h_\ell)]\Big]^{m}\right\}
\end{eqnarray}
The final result for the disordered-averaged free energy per spin is the 
difference of the two contributions (\ref{eq: energetic}), (\ref{eq: entropic}).
To determine the order parameter $m$ (not to be confused with the
magnetization) we must extremize the free energy with respect to
$m$. In the limit $n\to 0$, $m$ (which becomes a real number in $[0,1]$) may be
viewed as the probability of two configurations belonging to different pure states.

Finally, let us mention that one can easily recover the RS expressions
from the 1RSB framework. In particular,
the replica symmetry assumption corresponds to the system having one
pure state or ergodic component. This implies that the effective field
distribution at a given site becomes a delta function $P_{i}(h)=\delta(h-\tilde{h}_{i})$
and the order parameter is the distribution $W(\tilde{h})$ of the effective
fields. Equivalently, the functional $W[\{P\}]$ becomes:
\be
\label{eq: rs_W}
  \mathcal{W}^{RS}[\{P\}]=\int\!\mathrm{d}\tilde{h}\,W(\tilde{h})
       \prod_{h}\delta\Big[P(h)-\delta(h-\tilde{h})]
\ee
Similarly, for the right and left eigenvector functionals $U[\{\phi\}]$, $V[\{\psi\}]$
we have:
\begin{eqnarray}
\label{eq: rs_U}
  U^{RS}[\{\phi\}]=\int\!\mathrm{d}\tilde{x}\,\Phi(\tilde{x})
       \prod_{x}\, \delta\left[\phi(x)-\delta(x-\tilde{x})\right]
  \nonumber \\
  V^{RS}[\{\psi\}]=\int\!\mathrm{d}\tilde{y}\,\Psi(\tilde{y})
       \prod_{y}\,\delta\left[\psi(y)-\delta(y-\tilde{y})\right]
\end{eqnarray}
Upon inserting these replica-symmetric functionals in
(\ref{eq: U},\ref{eq: V},\ref{eq: funcW}) and setting $m=0$
(which also expresses the existence of one pure state), we find
that $W,\Phi,\Psi$ satisfy the triplet of replica symmetric equations
(\ref{eq: phi_rs},\ref{eq: psi_rs},\ref{eq: w_rs}). 

\subsection{Correspondence with cavity approach}
As expected, the assumptions invoked in the 1RSB setup
of both the cavity approach and the replica formalism describe exactly the
same physics of the ultrametric structure of the ergodic
components. In particular, for systems
with a Poisson distribution of long range connections at
a given site, the functional order parameters $U[\{\phi\}]$ and $W[\{P\}]$ 
of the replica formalism satisfy the equations of the cavity method
expressed in the functional form of (\ref{eq: functionalQ}, \ref{eq: functionalP}).
Upon using the identity 
\be
  \beta B(J,z)-\log 2\cosh(\beta z)=
  \log\frac{\cosh(\beta J)}{\cosh(\beta u(J,z))}
\ee
we may write
\be
\hspace*{-25mm}
  \rme^{\beta\mu\Delta F_{\sigma}}=
   \frac{\rme^{\mu\beta B(J_{0},x_{1})}}{[2\cosh(\beta x_{1})]^{\mu}}
   \prod_{\ell=1}^{k}\bigg\{
     \frac{\rme^{\mu\beta B(J_\ell/c,h_\ell)}}{[2\cosh(\beta h_\ell)]^{\mu}}\bigg\}
   \Big[2\cosh[\beta u(J_{0},x_{1})+\beta\sum_\ell u(J_\ell/c,h_\ell)]\Big]^{\mu}
\ee
After substituting the shorthand (\ref{eq: Qq}) into
the self-consistent equation (\ref{eq: U}) we see that with 
the simple correspondence: 
\bd
 \mu\to m \qquad \mathcal{F}[Q]\to U[\{\phi\}]
 \qquad C_{Q}^{-1}\to\lambda(m\vert \phi,M)
\ed
it becomes identical to (\ref{eq: functionalQ}).
To complete the correspondence between the two methods, 
we need also verify that $\mathcal{W}[\{P\}]$ given by (\ref{eq: funcW})
satisfies the second equation of the cavity method
(\ref{eq: functionalP}). To this end we will use
the 1RSB analogue of (\ref{eq: rs_xfy}) which relates the RS left- right
eigenvectors of the replicated transfer matrix. The appropriate generalization,
using the shorthand (\ref{eq: Qq}), can be written as:
\small
\begin{eqnarray}
\label{eq: 1rsb_xfy}
\hspace*{-25mm}
  U[\{\phi\}]=\int\!\{\rmd \psi\}\{\rmd
  M\}\,V[\{\psi\}]\mathcal{P}[\{M\}]\nonumber\\
\times \,
   \prod_{x}\delta\bigg[\phi(x)-\frac{[2\cosh(\beta x)]^{m}}{C_{x}(\psi,M)}
    \int\!\frac{\mathrm{d}y\mathrm{d}\theta\,\psi(y)M(\theta\vert m)}
             {[2\cosh(\beta y)]^{m}}\delta(x-y-\theta)\bigg]
\end{eqnarray}
\normalsize
with the normalization constant:
\be
  C_{x}(\psi,M)=\int\!\frac{\mathrm{d}y\mathrm{d}\theta\,\psi(y) M(\theta\vert m)}
                 {[2\cosh(\beta y)]^{m}}[2\cosh(\beta\theta+\beta y)]^{m}
\ee
Although here the algebra is more complicated, this can be shown in a
straightforward manner, similar to the RS case in section \ref{sec: RSlink},
by inserting the expression for $U[\{\phi\}]$ in (\ref{eq: U})
and then using (\ref{eq: V}). We next insert (\ref{eq: U}) and (\ref{eq: V})
in the right-hand side of (\ref{eq: funcW}) and use (\ref{eq: 1rsb_xfy})
to simplify the resulting expression, to find:
\be
\fl  \mathcal{W}[\{P\}]=\int\!\{\rmd \phi\}\{\rmd \phi^\prime\}\{\rmd M\}\,
    U[\{\phi\}]U[\{\phi^\prime\}]\mathcal{P}[\{M\}]
   \prod_{h}\,\delta\Big[P(h)-\mathcal{A}_{W}(h,\phi,\phi^\prime\vert M)\Big]
\ee
where
\begin{eqnarray}
\hspace*{-25mm}
  \mathcal{A}_{W}(h,\phi,\phi^\prime\vert M)=\frac{[2\cosh(\beta
  h)]^{m}}{C_{h}(\phi,\phi^\prime,M)} 
  \int\!\frac{
  \mathrm{d}x\mathrm{d}x^\prime\mathrm{d}\theta\,\phi(x)\phi(x^\prime)M(\theta\vert m)
   \rme^{\beta m B(J_{0},x)+\beta m B(J_{0},x^\prime)}}
        {[4\cosh(\beta x)\cosh(\beta x^\prime)]^{m}}
  \nonumber \\
\hspace*{60mm}\times\,
        \delta[h-\theta-u(J_{0},x)-u(J_{0},x^\prime)]
\end{eqnarray}
\begin{eqnarray}
C_{h}(\phi,\phi^\prime,M)=\int\!\frac{
  \mathrm{d}x\mathrm{d}x^\prime\mathrm{d}\theta\,\phi(x)\phi(x^\prime)M(\theta\vert m)
   \rme^{\beta m B(J_{0},x)+\beta m B(J_{0},x^\prime)}}
        {[4\cosh(\beta x)\cosh(\beta x^\prime)]^{m}}
  \nonumber \\
\hspace*{30mm}\times\,
  [2\cosh(\beta\theta+\beta u(J_{0},x)+\beta u(J_{0},x^\prime))]^{m}
\end{eqnarray}
With $\mathcal{P}[\{M\}]$ as in (\ref{eq: Qq}) this result is the cavity equation
(\ref{eq: functionalP}), since properties of the Poisson distribution
enable us to set in the latter 
$\sum_{k} k p_{k} f(k-1)/\bra k\ket\to\sum_{k}p_{k}f(k)$, 
and we also have
\begin{eqnarray}
  \rme^{\beta\mu\Delta F_{\tau}}=
   \frac{\rme^{\beta\mu B(J_{0},x_{1})+\beta\mu B(J_{0},x_{2})}}
        {[4\cosh(\beta x_{1})\cosh(\beta x_{2})]^{\mu}}
    \prod_{\ell=1}^{k}\left[\frac{\rme^{\beta\mu B(J_\ell/c,h_\ell)}}
                               {[2\cosh(\beta h_\ell)]^{\mu}}\right]
   \nonumber \\
\hspace*{20mm}\times
    \Big[2\cosh(\beta u(J_{0},x_{1})+\beta u(J_{0},x_{2})+
                \beta\sum_\ell u(J_\ell/c,h_\ell))\Big]^{\mu}
\end{eqnarray}
We conclude that for `small-world' models with Poissonian long range
connectivity, as expected,  the cavity fields are equivalent to the effective fields
introduced in the replica formalism. We note that this correspondence can 
be generalized to the case of an arbitrary connectivity distribution by 
exploiting the techniques used in \cite{WS87,LeVa} in the replica approach.

\section{Numerical results}
In \cite{Ni04} phase diagrams were displayed for Poissonian connectivity
distributions of average $c$ and long range interaction distributions 
$p_J(J) = p\delta(1-J) + (1-p)\delta(1+J)$ for various values of $p$ and $c$.
In line with physical reasoning \cite{PaTo} and experience with other spin glass 
models 
(e.g. the SK model \cite{SK}) it is assumed that
the spin glass to ferromagnetic boundary is parallel to the
temperature axis, i.e. it is 
inferred that no reentrance occurs for decreasing temperature. In the SK model,
this is correct only at the level of full RSB, whereas, for 
the replica symmetric
approximation one does find reentrance. Thus in RS there is a region of the
spin glass phase which in reality (in full RSB theory and simulations)
corresponds to a mixed phase with non-zero magnetization, implying a clear 
difference between RS and RSB results. 
To test our 1RSB theory, we have applied it to a model similar to the one in
\cite{Ni04}, but with fixed connectivity.
Unfortunately the fluctuations in the population dynamics 
algorithms we use (for a tolerable CPU cost) caused by a
non-peaked connectivity distribution are found to severely limit the 
accessible accuracy in the value of $\mu$.

\begin{figure}[t]
\begin{picture}(80,150)
\put(100,10){\includegraphics[height=5.0cm,width=6.0cm]{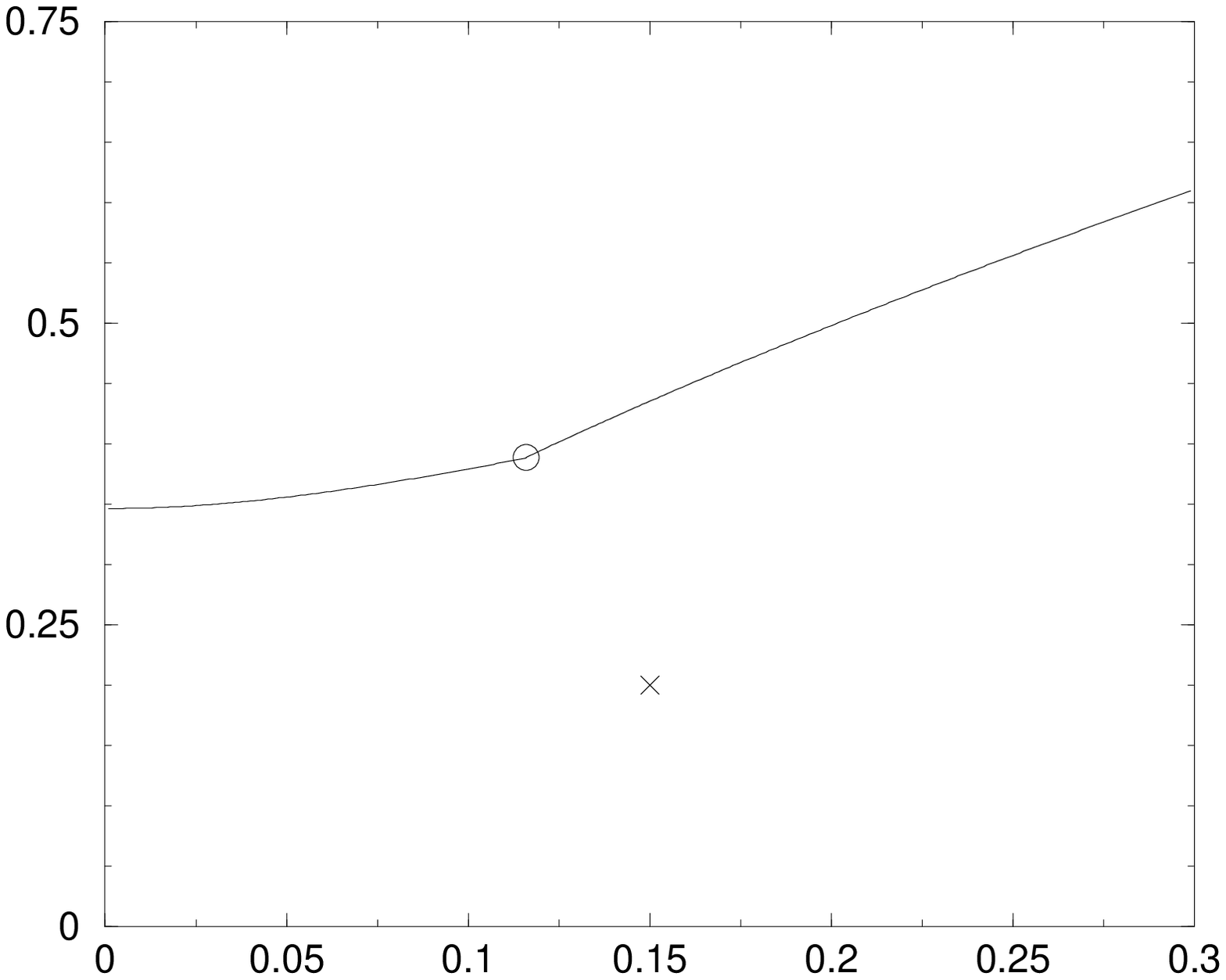}}
\put(80,70){\small $T$}
\put(180,0){\small $J_0$}
\put(150, 120){\small $P$}
\put(130, 50){\small $SG$}
\put(210, 50){\small $F$}
\end{picture}
\caption{Phase diagram, with left to the triple point (circle) the 
P $\to$ SG transition line and to its right hand side the P $\to$ F 
transition. Our population dynamics procedure concentrates on the point 
marked by a cross}
\label{fig:phasediag}
\end{figure}

The P $\to$ F and P $\to$ SG phase transition lines in the RS phase 
diagram for the model with $p(k) = \delta_{k,6}$, 
$p_J(J) = \frac{5}{8}\delta(J-1) + \frac{3}{8}(J+1)$ are found by solving 
 equations (\ref{eq:fbif}) and (\ref{eq:sgbif}) numerically. 
We have plotted the phase diagram in figure \ref{fig:phasediag}. 

A non-reentrant phase diagram would suggest that the SG $\to$ F transition
line runs vertically down from the triple point, so would be fully
determined by the corresponding value for $J_0$. We have attempted to
find support for this 
conjecture, by concentrating on a point within the ordered region of
the phase diagram, which 
RS population dynamics predicts is just outside the ferromagnetic 
phase. If for this point $J_0$ is larger than the $J_0$ corresponding to 
the triple point, one expects a nonzero magnetization at a certain level of 
RSB.

\begin{figure}[t]
\begin{picture}(80,150)
\put(20,10){\includegraphics[height=5.0cm,width=5.0cm]{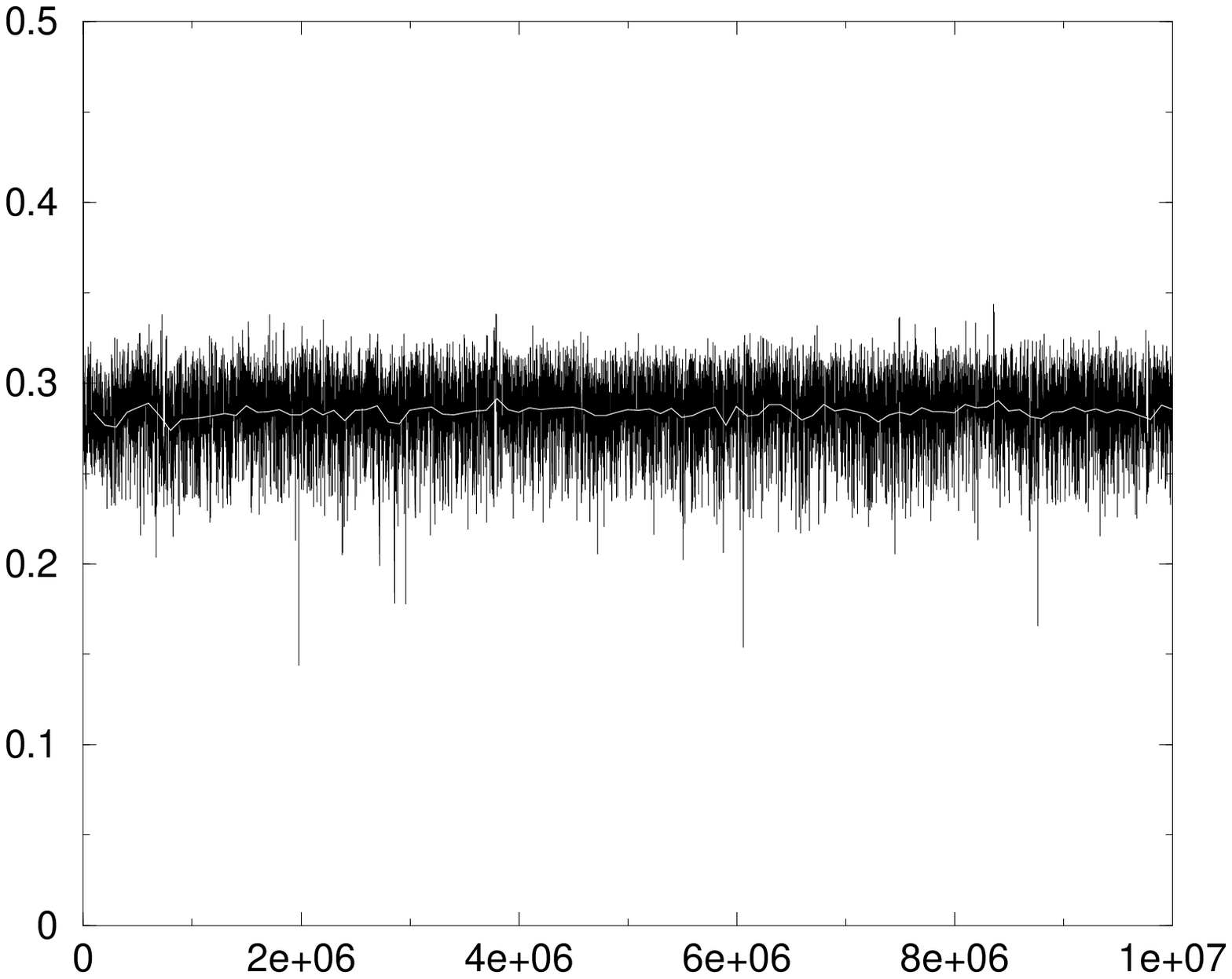}}
\put(160,10){\includegraphics[height=5.0cm,width=5.0cm]{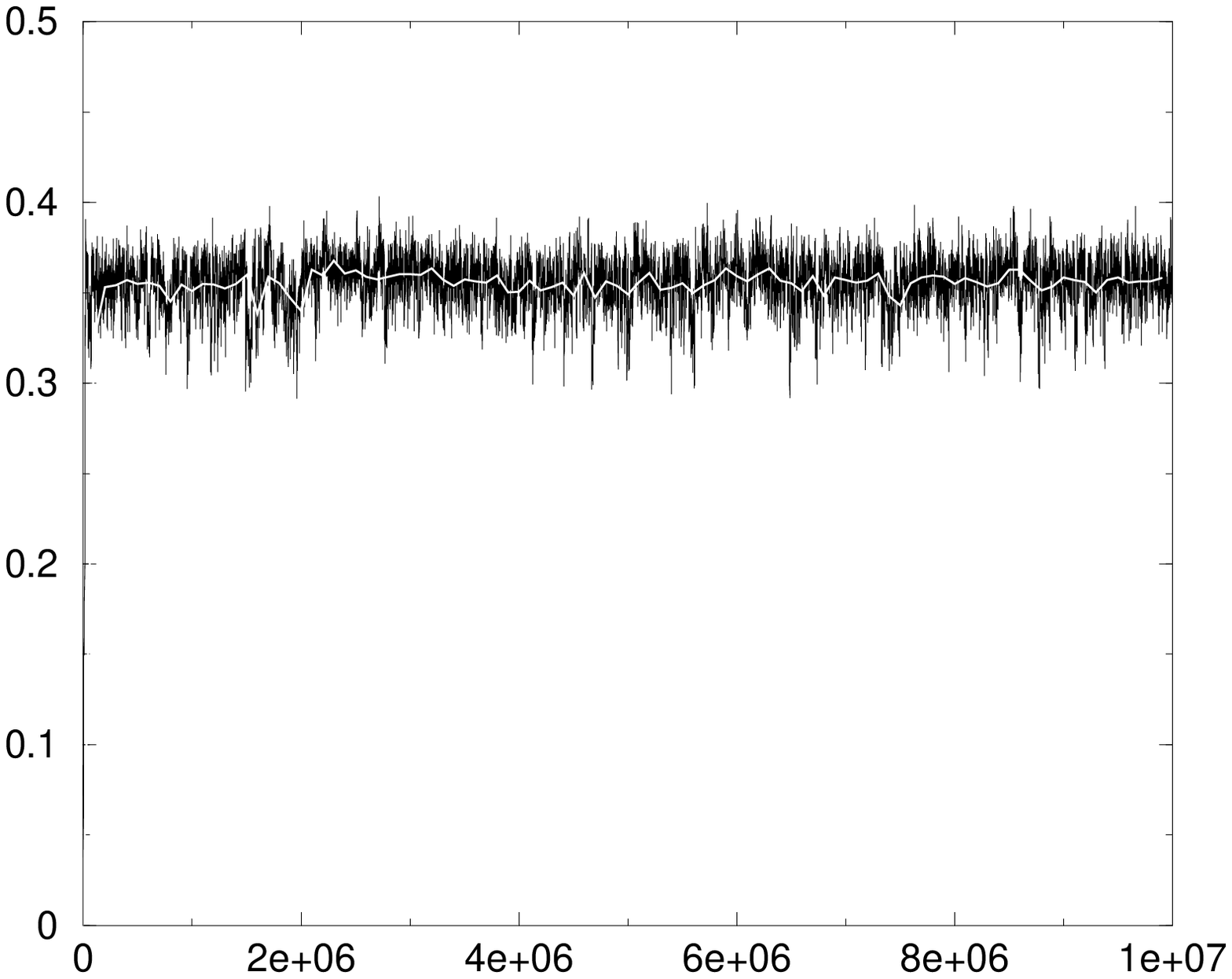}}
\put(300,10){\includegraphics[height=5.0cm,width=5.0cm]{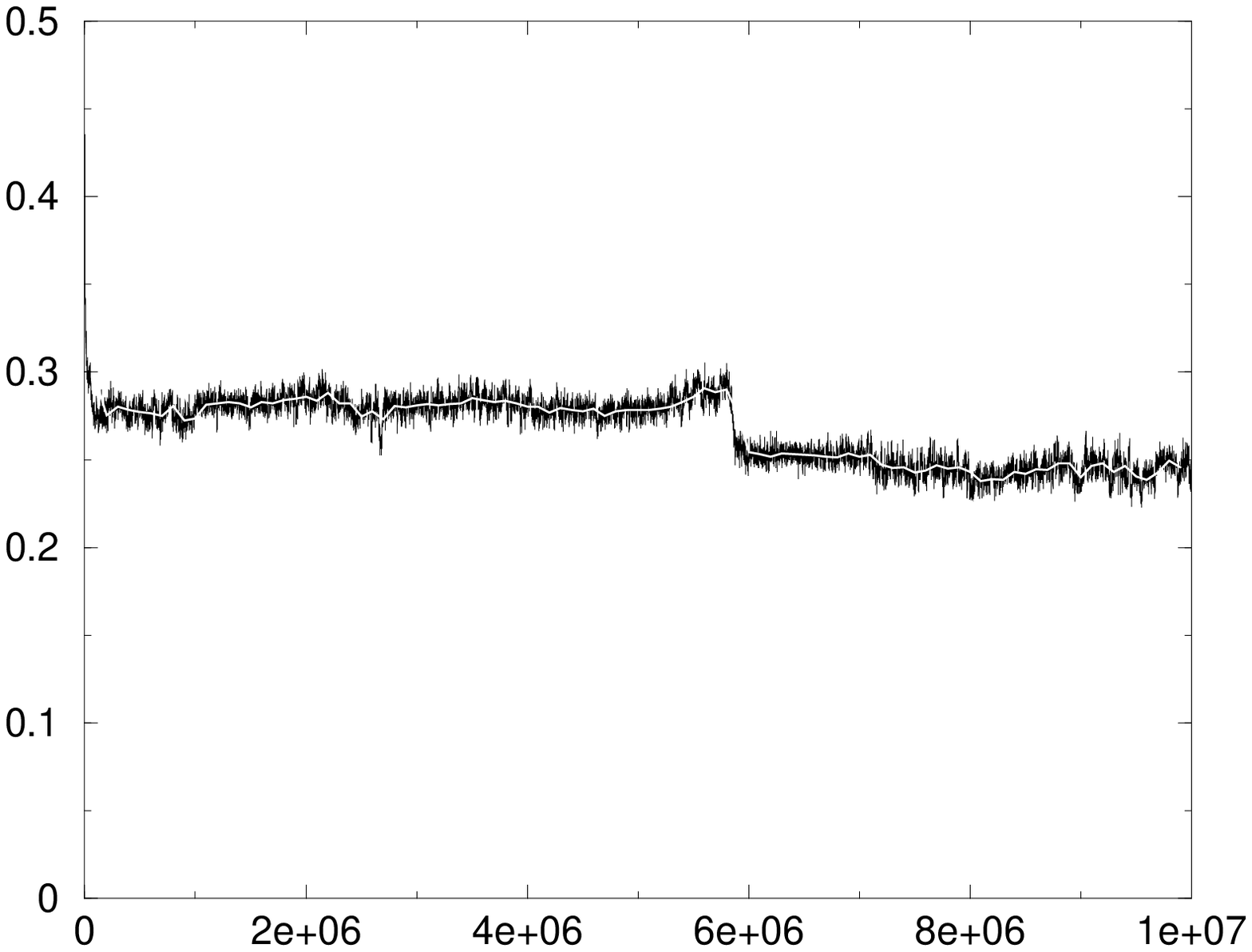}}
\put(5,80){\small $m$}
\put(90,0){\small $t$}
\put(230,0){\small $t$}
\put(370,0){\small $t$}
\end{picture}
\caption{Left to right: we plot the magnetization ($m$) against the
  number of updates per spin in simulations with $N = 2000$, $N = 10\,000$ and
  $N = 50\,000$ spins. The black lines show the magnetisations averaged
  over 10 observations, with one observation every 50 updates per
  spin. The white
  lines are averages over 2000 observations. As expected the
  fluctuations in the magnetization decrease with system size. Note
  the large change in the magnetisation for $N = 50\,000$ system after
  $\approx 6 \times 10^6$ updates per spin, suggesting that the system
  has not equilibrated yet.
}
\label{simulations}
\end{figure}

For $T=0.2$, the maximal value of $J_0$ 
for which the RS population dynamics algorithm predicts a zero magnetization
is approximately $J_0=0.15$. Corresponding values of other observables
are
\begin{equation}
f=-0.3561 \qquad \mbox{ and } \qquad q=0.5789
\end{equation}
The triple point is at $J_0=0.116$, and thus the
correct magnetization is not expected to vanish.

Running the 1RSB population dynamics algorithm at this point in the phase 
diagram for different $\mu$, we found that for $\mu = (0.31, 0.32,
0.33)$ the values of $\rmd F/\rmd \mu$ are $(3.7\times
10^{-5}, -0.5 \times 10^{-5}, -1.3 \times 10^{-5})$ respectively. We conclude
that the correct value for $\mu$ (where $\rmd F/\rmd \mu$
intersects the zero-axis) is $\mu = 0.32 \pm 0.01$.
For a system of $\cN = 2000$ and $\cM = 1000$ we find the results
\begin{eqnarray}
f=-0.3557 \pm 0.0001 \qquad \mbox{ and } \qquad q_0 = 0.397 \pm
0.003\nonumber \\
q_1 = 0.673 \pm 0.003 \qquad \hspace{6.5mm} \mbox{    and } \qquad m =
0.2 \pm 0.05 
\end{eqnarray}
i.e., although the fluctuations in the magnetization are large, it
is clearly nonzero. 

Further support for a nonzero magnetization is 
given by simulation results: in figure \ref{simulations} we show
simulation results for three values of the  
system size ($N=2000$, $N = 10,000$ and $N=50,000$), where we plot the
magnetization versus the number of updates per spin up to $10^7$
sequential Glauber updates. The results show 
strong finite size effects coupled with slow relaxation towards
equilibrium, the latter point precluding simulations of larger
systems. To check that the magnetizations were not just the result of
very slow relaxation towards equilibrium, we ran simulations of an
$N=2000$ system for $50 \times 10^6$ updates per spin, and we started other
simulations with small initial magnetization. In both cases we 
observed that the resulting long time magnetization was still $\approx
0.25$. The average magnetisation and the
standard deviation $(\Delta m)$ over 10 independent runs (different
realisation of both the thermal noise and disorder) gives
\begin{eqnarray}
N=10\,000: \qquad m_{\rm av} = 0.24 && \hspace*{25mm}\Delta m = 0.07 \\
N=50\,000: \qquad m_{\rm av} = 0.25 && \hspace*{25mm}\Delta m = 0.03 
\end{eqnarray}
Thus we believe that the fluctuations within runs and between runs can
both be put down to finite size effects.
Although from the simulations performed so far 
the equilibrium value of the magnetisation in the thermodynamic limit
cannot be determined with great accuracy, 
the 1RSB average value of $m=0.2$ certainly seems more acceptable than the
vanishing magnetisation result of RS theory.

\section{Conclusions}

In this paper we have applied the cavity approach to the `small world'
lattice model, previously studied using replica theory in \cite{Ni04}.
The model is generalised to the case of an arbitrary (site independent)
connectivity distribution for the random graph component of the lattice.
Along the lines of \cite{MePa01}, we have extended the RS results to the
case of a 1RSB formalism, which is numerically solvable with population dynamics
algorithms. Furthermore, we have extended the replica formalism (based on
replicated transfer matrices) at the level of 1RSB
according to the Parisi-scheme and shown
how it recovers exactly the functional order parameter equations 
of the cavity method.
We applied the population dynamics algorithm to a model in which the
random graph component has a connectivity distribution 
$p(k) = \delta_{k,6}$. We find support for the conjecture that the 
replica symmetric reentrance in the spin glass phase is non-physical:
the ferromagnetic to spin glass phase boundary in 1RSB shifts into the
spin glass region of the RS phase diagram. This result is supported
by simulations.

Possible extensions to this approach include examining the case where
one has more neighbours along the spin chain, e.g. one could include
next-nearest neighbour interactions which would increase the
clustering effect in one dimension. This would lead to order
parameters encompassing the joint distribution of effective
fields. We are also investigating a model of $XY$ spins on a `small
world' architecture, the continuous nature of the spins leads to
interesting new behaviour and requires new techniques. 

Further stages of replica symmetry breaking although in
principle attainable would of course have a prohibitive CPU cost.
The exception is possibly at zero temperature, 
where expressions simplify and analysis
can be pushed further, along the lines of recent papers as
\cite{CastellaniRicci,MontanariParisiRicci,Rizzo}. 
There are of course many other
disordered problems where the small world architecture is appropriate
and the techniques presented here could be applicable.

\section{Acknowledgments}
The authors are indebted to I. P\'erez Castillo, A.C.C. Coolen and
N.S. Skantzos for illuminating discussions on finite connectivity issues. 
TN and BW acknowledge financial support from the State Scholarships
Foundation (Greece) and the FOM Foundation
(Fundamenteel Onderzoek der Materie, the Netherlands),
respectively.

\end{document}